\documentclass[12pt,preprint]{aastex}

\shorttitle{Runaway Microquasar LS~5039}
\shortauthors{McSwain et al.}

\received{2003 July 3}
\accepted{2003 September 23}
\begin{document}

\title{The N Enrichment and Supernova Ejection of the Runaway Microquasar 
LS~5039\footnote{Based on observations with the NASA/ESA Hubble Space Telescope
obtained at the Space Telescope Science Institute, which is operated by the 
Association of Universities for Research in Astronomy, Incorporated, under NASA
contract NAS5-26555.}}

\author{M. V. McSwain\altaffilmark{2},
        D. R. Gies\altaffilmark{3}, 
        W. Huang\altaffilmark{3,4},
        P. J. Wiita, D. W. Wingert}

\affil{Center for High Angular Resolution Astronomy\\
Department of Physics and Astronomy \\
Georgia State University, University Plaza, Atlanta, GA  30303-3083\\
Electronic mail: mcswain@chara.gsu.edu, gies@chara.gsu.edu,
huang@chara.gsu.edu, wiita@chara.gsu.edu, wingert@chara.gsu.edu}

\altaffiltext{2}{Visiting Astronomer, Cerro Tololo Interamerican Observatory,
National Optical Astronomy Observatories, operated by the Association
of Universities for Research in Astronomy, Inc., under contract with
the National Science Foundation.}

\altaffiltext{3}{Visiting Astronomer, University of Texas McDonald Observatory.}

\altaffiltext{4}{Visiting Astronomer, Kitt Peak National Observatory,
National Optical Astronomy Observatories, operated by the Association
of Universities for Research in Astronomy, Inc., under contract with
the National Science Foundation.}

\author{L. Kaper}
\affil{Astronomical Institute ``Anton Pannekoek'', University of Amsterdam,
Kruislaan 403, NL-1098 SJ Amsterdam, The Netherlands\\
Electronic mail: lexk@science.uva.nl}

\slugcomment{2004, ApJ, 600, Jan. 10 issue, in press.}
\paperid{58557}


\begin{abstract}
We present an investigation of new optical and ultraviolet spectra
of the mass donor star in the massive X-ray binary LS~5039. 
The optical band spectral line strengths 
indicate that the atmosphere is N-rich and C-poor, 
and we classify the stellar spectrum as type ON6.5~V((f)).   
The N-strong and C-weak pattern is also found in the stellar 
wind P~Cygni lines of \ion{N}{5} $\lambda 1240$ and \ion{C}{4} $\lambda 1550$
(narrow absorption components in the former indicate that
the wind terminal velocity is $V_\infty = 2440 \pm 190$ km~s$^{-1}$).
We suggest that the N-enrichment may result 
from internal mixing if the O-star was born as a rapid rotator, 
or the O-star may have accreted N-rich gas prior to a 
common-envelope interaction with the progenitor of the supernova.   
We re-evaluated the orbital elements to find an orbital period of 
$P=4.4267 \pm 0.0005$~d.  We compared the spectral 
line profiles with new non-LTE, line-blanketed model spectra 
from \citet{lan03}, from which we derive an effective temperature
$T_{\rm eff} = 37.5 \pm 1.7$~kK, gravity $\log g = 4.0 \pm 0.1$, 
and projected rotational velocity $V\sin i = 140 \pm 8$ km~s$^{-1}$.
We fit the UV, optical, and IR flux distribution using a model 
spectrum and extinction law with parameters $E(B-V)= 1.28 \pm 0.02$
and $R= 3.18 \pm 0.07$.   We confirm the co-variability of the 
observed X-ray flux and stellar wind mass loss rate derived from 
the H$\alpha$ profile \citep{rei03}, which supports the wind accretion 
scenario for the X-ray production in LS~5039.   
Wind accretion models indicate that the compact companion has a mass 
$M_X/M_\odot = 1.4\pm0.4$, consistent with its identification as a neutron star.
We argue that the O-star has a mass in the range 20 -- $35 M_\odot$ 
(based upon a lower limit for the distance and the lack of eclipses). 
The observed eccentricity and runaway velocity of the binary 
can only be reconciled if the neutron star received a modest 
kick velocity due to a slight asymmetry in the supernova explosion
(during which $>5 ~M_\odot$ was ejected).    
\end{abstract}



\section{Introduction}                              

Massive stars in close binaries are destined to interact 
in some way over their lifetimes.  The initially more massive 
star is the first to evolve, and after shedding its outer, 
H-rich envelope by stellar winds, it may begin mass transfer 
of CNO nuclear processed gas to its companion \citep{del94}.  
After completion of the mass transfer stage, the abundance 
ratio of N/C in the photosphere of the gainer may have increased 
from the solar value of 0.3 to an enriched value of 3 \citep{deg92},
or even higher if the mass accretion had induced large scale mixing in 
the mass gainer \citep{van94}.   However, there is conflicting evidence
about whether or not mass transfer actually occurs in real binaries 
composed of massive, O-type stars \citep{lwp03}.   The massive 
X-ray binaries (MXRBs), for example, generally do not exhibit 
evidence of N-enrichment, and this fact suggests that the supernova (SN) 
progenitor may eject mass from the entire system, perhaps during 
a Luminous Blue Variable stage \citep{dea77,wel99}.   

We were surprised, therefore, to discover that the \ion{C}{4} 
$\lambda\lambda 5801,5812$ lines in the spectrum of the 
MXRB and microquasar LS~5039 (type O6.5~V((f)); \citet{cla01})
displayed a weakness indicative of CNO processed gas \citep{mcs01}.   
LS~5039 is also unusual in several other respects.  
It is one of only a few confirmed MXRBs
with associated radio emission \citep{rib99,par02}.
It has radio-emitting relativistic
jets characteristic of galactic microquasars, and
it is probably a high energy gamma ray source as well \citep{par00}.
We found that the system is a short period binary 
with the highest known eccentricity ($e = 0.41\pm0.05$)
among O star binaries with comparable periods \citep{mcs01}.
This high eccentricity probably resulted from the huge mass
loss that occurred with the SN explosion that gave
birth to the compact star in the system \citep{bha91,nel99}.
Binaries that suffer large mass loss in a SN are expected
to become runaway stars, and recently both we \citep{mcs02} and
\citet{rib02} found that LS~5039 has a significant proper
motion indicating a large peculiar space velocity.  

Here we present results from new optical and ultraviolet 
spectroscopy that confirm the presence of CNO processed 
gas in the photosphere of LS~5039.  We give revised orbital 
elements for the binary (\S2), and we use our combined 
optical spectra to assign LS~5039 to the category of the 
N-rich ON stars (\S3).   We present the first UV spectra 
of LS~5039 (made with HST/STIS) that we analyze to determine 
the stellar wind properties (\S4) and interstellar 
reddening (\S5).  We use the new results to revisit the 
issue of the masses of the O-star and compact companion (\S6). 
We also present simple models for the change in orbital 
elements caused by the supernova (\S7), which require a  
modest kick velocity in order to reproduce the 
observed eccentricity and runaway velocity.   
Finally we discuss the possible explanations for 
the N-enrichment given our current understanding of 
the evolutionary history of the system (\S8).    


\section{Optical Spectroscopy and Orbital Elements} 

We obtained new spectra of LS~5039 on four different observing 
runs, and the details about the spectrograph setup in each case
are summarized in Table~1.  The primary set of blue region 
spectra come from observations with the University of Texas 
McDonald Observatory 2.7~m telescope and Large Cassegrain 
Spectrograph (LCS) on two observing runs (see \citet{gie02} for details).
All these were made with a TI CCD ($800\times800$, 
15.2 $\mu$m pixels) except on one night when we used a 
Loral/Fairchild CCD ($1024\times1024$, 12 $\mu$m pixels).
Most of the spectra were made in a single 1200~s exposure
that resulted in a signal-to-noise ratio of S/N = 120 in the continuum.  
A secondary set of blue spectra were made with the CTIO
1.5~m telescope and Cassegrain spectrograph (CS) using a Loral 
CCD ($1024\times1024$, 15 $\mu$m pixels).  These were 
usually 1800~s exposures that produced a S/N = 80 in the continuum.  
We also observed the red spectral region during a
KPNO Coude Feed run in 2002 June.  These spectra are comparable to 
those we presented in \citet{mcs01} except that these
new spectra were obtained 
with a TI CCD that had smaller wavelength coverage.  
Finally, we obtained one additional red spectrum of lower resolving power 
with the CTIO 1.5~m CS on 2003 March 21.
All these spectra were extracted and calibrated using standard techniques in 
IRAF\footnote{IRAF is distributed by the
National Optical Astronomy Observatories, which is operated by
the Association of Universities for Research in Astronomy, Inc.,
under cooperative agreement with the National Science Foundation.}.
We removed atmospheric telluric lines from the red spectra using 
spectra of broad-lined A-type stars (see \citet{mcs01}).  

\placetable{tab1}      

Our first task was to re-evaluate the orbital elements 
presented in \citet{mcs01} using an enlarged set of radial velocity 
measurements.   We measured radial velocities for the blue spectra 
by cross-correlating each spectrum with the average of the 
blue spectra made at McDonald Observatory (the largest homogeneous set). 
We omitted spectral regions containing interstellar features. 
These relative velocities were converted to an absolute scale 
by adding the radial velocity of the average spectrum as 
determined by parabolic fitting of the cores of the strong lines, 
$+11.2 \pm 9.2$ km~s$^{-1}$ (using rest wavelengths from 
\citet{bln78}).   Radial velocities from the red spectra were also  
determined by cross-correlation techniques, but in this case the 
reference spectrum was formed by a shift-and-add average of the 
red spectra obtained earlier \citep{mcs01}.  This cross-correlation function 
is generally based on the positions of the H$\alpha$ and \ion{He}{1} $\lambda 6678$ 
features, but for the CTIO red spectrum we also included the spectral 
regions containing \ion{He}{1} $\lambda 5876$ and \ion{He}{2} $\lambda 6527$.
We also determined the cross-correlation 
positions of the sharp interstellar line at 6614 \AA ~that we used to 
make small velocity corrections between observatories. 
The 14 new radial velocity measurements are collected in Table~2. 

\placetable{tab2}      

Our first attempt to revise the orbital period and other elements 
starting with the period we found earlier ($P=4.117$ d; \citet{mcs01})
failed, with 8 of the 14 new measurements falling far from the 
calculated radial velocity curve.  Thus, we made a new period 
search using the discrete Fourier transform method of \citet{rob87}. 
There are a number of candidate peaks in the periodogram that are clustered 
around a frequency of 0.25 cycles~d$^{-1}$, and most of these 
are aliases introduced by our sparse
data sampling \citep{mcs01}.  We tested each of these candidate periods 
using the nonlinear, least-squares fitting program of \citet{mor74} to 
determine the orbital elements, and only one of the test periods, 
$P=4.4267 \pm 0.0005$~d, led to consistent placement of all the new measurements. 
This value was also the primary signal found using the Phase 
Dispersion Minimization technique of \citet{ste78} that is useful for 
finding periodicities in non-sinusoidal variations.  
Significantly, this period also fits new radial velocity data obtained 
by J.\ Casares,  M.\ Rib\'{o}, J.\ M.\ Paredes, and J.\ Mart\'{\i} (in preparation). 
The revised orbital elements are given in Table~3, and 
the observations and radial velocity curve are shown in Figure~1.  
This fit assumed equal weights for each radial velocity measurement;
fits weighted using the inverse square of the standard deviation 
of each measurement led to identical results within errors. 
Aside from the orbital period, the new elements do not differ greatly from 
those we presented earlier \citep{mcs01}.  Most of our observations are clustered 
into runs of a week or so (comparable to the period) so that while the
placement of these groups of measurements in phase is somewhat altered, 
the overall shape of the curve is much the same. 

\placetable{tab3}      

\begin{figure}
\rotatebox{90}{
\epsscale{0.7}
\plotone{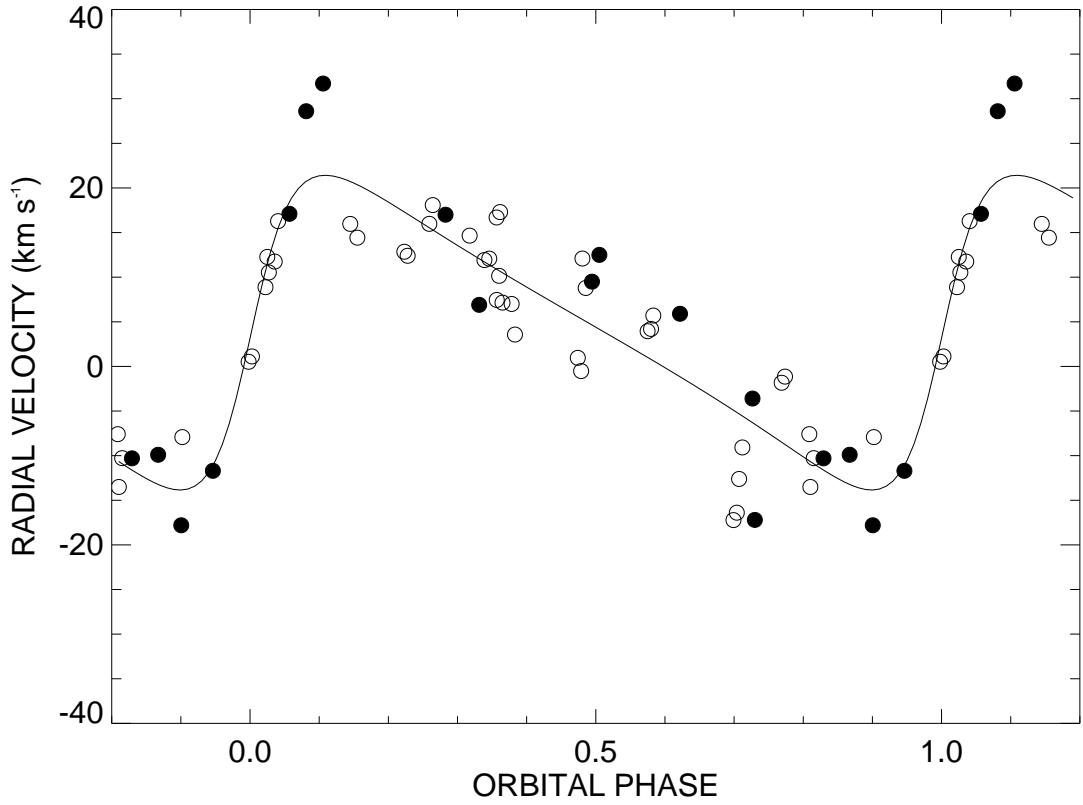}}
\caption{The revised radial velocity curve ({\it solid line}) 
for LS~5039 together with the original measurements from 
\citet{mcs01} ({\it open circles}) and the new measurements
({\it filled circles}).}
\label{fig1}
\end{figure}

\placefigure{fig1}     


\section{Spectral Classification and Stellar Parameters} 

\citet{cla01} discuss the appearance of the optical and near-IR 
spectrum of LS~5039, from which they derive a spectral 
classification of O6.5~V((f)) for the mass donor star. 
Here we examine the line patterns in our optical spectra 
to confirm this classification.  We formed an average blue 
spectrum by calculating a shift-and-add mean spectrum 
from our McDonald and best CTIO spectra (based upon the 
line shifts from the orbital solution in the last section). 
The average interstellar spectrum was divided out of each 
observed spectrum prior to forming the average.  The resulting 
mean spectrum is shown in the upper plot in Figure~2 
together with identifications for the more prominent lines. 
We also show in this diagram the blue spectra of the standard 
stars HD~93146 (O6.5~V((f)); from the atlas of \citet{wal90}) 
and 15~Mon (O7~V((f)); obtained during our CTIO run).  
\citet{lan03} have recently made available a library of 
synthetic stellar spectra for a grid of non-LTE atmospheres
appropriate for O-type stars\footnote{http://tlusty.gsfc.nasa.gov}, 
and we show in Figure~2 a flux rectified version of one of their models 
(with stellar parameters discussed below).  This spectrum was 
convolved with broadening functions to simulate stellar rotational 
and spectrographic instrumental broadening (discussed below).  

\begin{figure}
\rotatebox{90}{
\epsscale{0.7}
\plotone{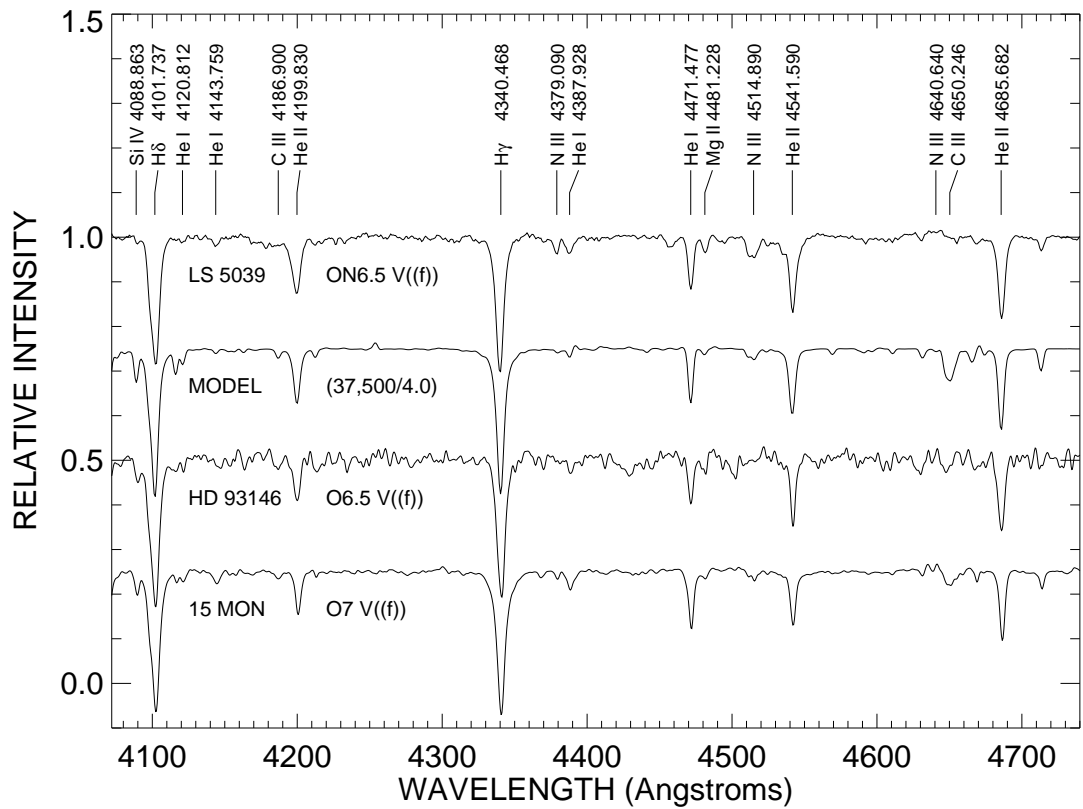}}
\caption{The average spectrum of LS~5039 compared to the model 
spectrum from \citet{lan03}, an O6.5~V((f)) standard from 
\citet{wal90}, and an O7~V((f)) standard (from our CTIO run).
Some of the stronger lines are identified at the top. 
The lines of \ion{N}{3} are systematically stronger 
(and the \ion{C}{3} lines weaker) in 
the spectrum of LS~5039 than in the other spectra.}
\label{fig2}
\end{figure}

\placefigure{fig2}     

The spectral subtype among O-type stars is defined by the relative 
strengths of the \ion{He}{2} and \ion{He}{1} lines \citep{wal90}. 
The relative strengths of the \ion{He}{2} $\lambda 4541$ and
\ion{He}{1} $\lambda 4471$ lines in the spectrum of LS~5039 
make a good match with the O6.5~V((f)) standard star (but not 
with the cooler O7~V((f)) spectrum in which \ion{He}{1} $\lambda 4471$
is stronger).  The primary luminosity criterion at this spectral 
type is the strength of the \ion{He}{2} $\lambda 4686$ line 
that ranges from strong absorption to strong emission between 
main sequence and luminous supergiant stars (see Fig.~9 in \citet{wal90}). 
The deep \ion{He}{2} $\lambda 4686$ absorption line in the spectrum 
of LS~5039 clearly indicates a luminosity class V object.  
We also observe weak emission in the \ion{N}{3} $\lambda\lambda 
4634-4640-4642$ blend, which is indicated by the classification
suffix ((f)).  Our results therefore confirm the spectral 
classification obtained by \citet{cla01} of O6.5~V((f)). 

We note, however, the unusual strength of the \ion{N}{3} $\lambda 4379$ 
and \ion{N}{3} $\lambda 4510-4514-4518$ features in the spectrum 
of LS~5039 compared to the other reference spectra.  These lines 
tend to peak near type O8.5, and at any given subtype they increase
in strength with luminosity \citep{mat89}.   The equivalent width 
of the \ion{N}{3} $\lambda 4510-4514-4518$ blend in the spectrum 
of LS~5039 is $W_\lambda = 0.40 \pm 0.04$\AA , which is larger than that 
in all 13 N-strong stars with spectral types O7 or earlier listed by 
\citet{mat89} (with the exception of the N enhanced, O7~V((f)) star,
HD~90273).  Furthermore, we observe that the \ion{C}{3} $\lambda 4650$ 
blend is weak or absent in the spectrum of LS~5039, and other C 
features are also unusually weak (\ion{C}{3} $\lambda 4187$, 
\ion{C}{3} $\lambda 5696$ emission, \ion{C}{4} $\lambda 5801, 5812$). 
We show below that this trend of strong N and weak C is also 
found in the UV stellar wind lines (\S5).  These characteristics
are the hallmark of the ON type stars that appear to be contaminated
with CNO nuclear processed gas \citep{wal76,bln78,sch88,mat89}. 
One of the hottest stars that is currently assigned to the ON group is HD~110360 
(ON7~V; \citet{mat89}), and the strong similarity in appearance 
between LS~5039 and HD~110360 (see Fig.~5 in \citet{mat89}) suggests
that LS~5039 also belongs to this group.  Thus, we advocate a 
classification of ON6.5~V((f)) for the spectrum of LS~5039. 

The appearance of the optical line spectrum can also be used to 
help estimate the physical properties of the star using the 
new non-LTE model spectra from \citet{lan03}.   A direct comparison of 
their grid of model spectra with our observed spectrum can 
provide reliable estimates of the stellar projected rotational 
velocity $V\sin i$, the effective temperature $T_{\rm eff}$, 
and the gravitational acceleration $\log g$.   
We adopted the \citet{lan03} solar metallicity models, which 
assume an atmospheric microturbulent velocity of 10 km~s$^{-1}$.   
We used our relatively higher dispersion red spectra 
($\lambda/\Delta\lambda = 3510$; \citet{mcs01}) to 
find a value of $V\sin i$ that would produce a match  
of the model and observed profiles of the strong \ion{He}{1} $\lambda 5876$ 
line.  We formed a rotational broadening function for a linear 
limb darkening law \citep{gra92} using a limb darkening coefficient
from the tables of \citet{wad85}, and then the \citet{lan03}
model spectrum was convolved with this and a Gaussian instrumental 
broadening function for direct comparison with the observed profile. 
The best fit was made with $V\sin i = 140 \pm 8$ km~s$^{-1}$, 
which is in good agreement with our earlier result of 
$V\sin i = 131 \pm 6$ km~s$^{-1}$ based upon the half-width of the 
\ion{O}{3} $\lambda 5592$ line \citep{mcs01}.  

We estimated the effective temperature by comparing the observed
\ion{He}{2} to \ion{He}{1} line ratios with those predicted from 
the grid of model spectra from \citet{lan03}.  We measured 
equivalent widths of the major He lines by Gaussian fitting of 
the observed spectral profiles, and we then formed logarithms of the
important ratios used in spectral classification (Table~4).  
We made numerical integrations across the same lines in the 
grid of model spectra by \citet{lan03} and then formed plots 
of $\log [W_\lambda($\ion{He}{2}$) / W_\lambda($\ion{He}{1}$)]$ 
versus $T_{\rm eff}$.  We restricted the grid to models with 
$\log g = 4.0$, which is the appropriate gravity for main sequence 
stars (see below).   Table~4 lists the resulting values of $T_{\rm eff}$
made by interpolation with the observed ratios in these curves. 
We formed a final mean temperature by double weighting the 
result from the primary ratio that defines the classification sequence, 
\ion{He}{2} $\lambda 4541$ to \ion{He}{1} $\lambda 4471$, 
to arrive at $T_{\rm eff} = 37.5\pm1.7$~kK.   This temperature 
is somewhat low for spectral type -- temperature calibrations based 
upon pure H and He model atmospheres \citep{vac96} but it agrees well 
with more recent calibrations that rely on blanketed, non-LTE models 
\citep{mar02}.  

\placetable{tab4}      

The stellar gravity, $\log g$, is presumably the key parameter defining the 
appearance of the \ion{He}{2} $\lambda 4686$ line for stars of this 
temperature \citep{wal90}.  However, the emergence of this feature as an 
emission line in more luminous stars is the result of their vigorous 
stellar winds, and unfortunately the \citet{lan03} models are based 
on static, plane parallel atmospheres that will not predict the 
\ion{He}{2} $\lambda 4686$ line strength correctly in most cases. 
Instead, we can use the Stark broadening of the H Balmer lines to 
help estimate the atmospheric pressure and hence gravity.  H$\gamma$ 
is the most suitable feature in our spectra for this purpose since
H$\delta$ is severely blended with \ion{N}{3} 
$\lambda 4097$ at this spectral resolution.   We show in 
Figure~3 the observed and model H$\gamma$ profiles for 
$\log g = 3.7$, 4.0, and 4.3 (cgs units).   These model profiles
were convolved with both the rotational and instrumental 
broadening functions.  The comparison 
indicates that $\log g = 4.0\pm0.1$.  However, we caution that this 
value may be biased upwards by the presence of \ion{N}{3} lines in the 
wings of H$\gamma$ (wavelengths from the NIST 
database\footnote{http://physics.nist.gov/cgi-bin/AtData/main\_asd})
that are probably underestimated by the 
\citet{lan03} models in this particular case.   

\begin{figure}
\rotatebox{90}{
\epsscale{0.7}
\plotone{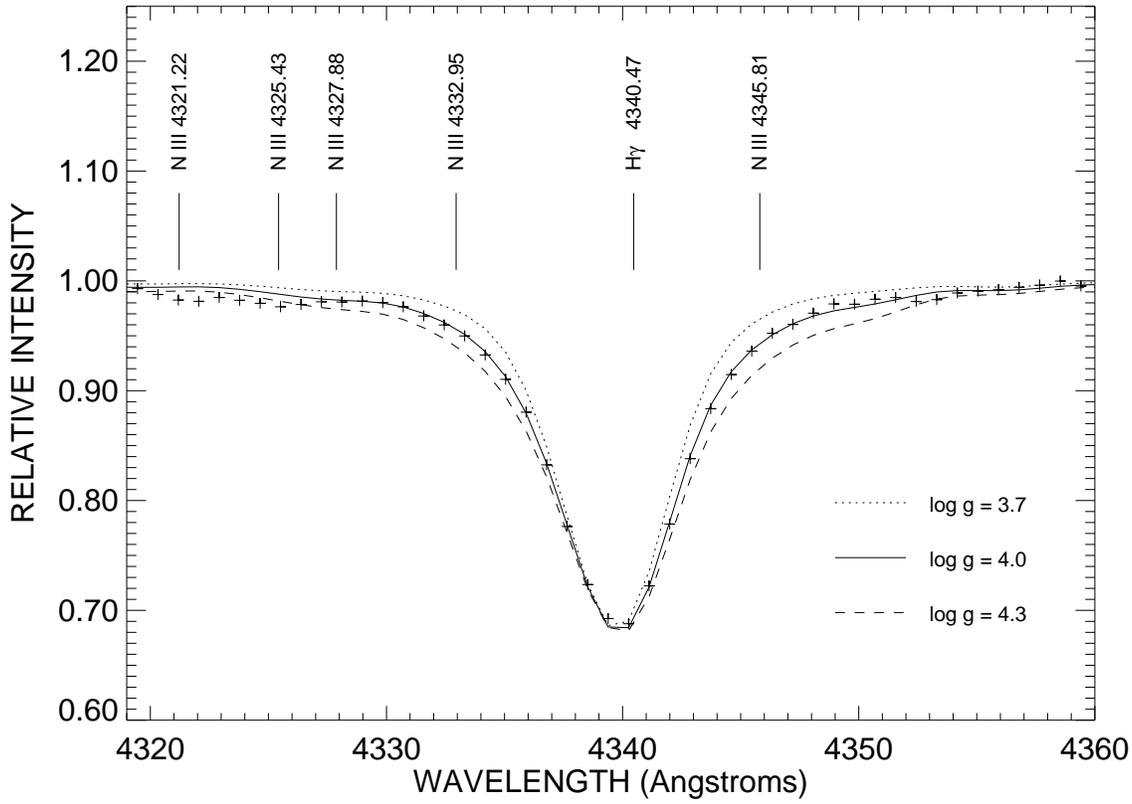}}
\caption{The H$\gamma$ profile in the average spectrum of LS~5039 
({\it plus signs}) compared to the model profiles from \citet{lan03} 
for three values of atmospheric gravity (and corresponding pressure
broadening of the Balmer line wings).  Possible line blends from 
transitions of \ion{N}{3} are identified above.}
\label{fig3}
\end{figure}

\placefigure{fig3}     
  
A reliable abundance analysis is clearly desirable, but will 
require higher dispersion spectra to account for the effects 
of line blending in the vicinity of the key CNO lines.  
However, we made preliminary estimates by calculating line 
profiles for a range of assumed abundances using the radiative transfer 
code {\it Synspec} and the non-LTE stellar atmosphere associated with 
our derived values of $T_{\rm eff}$ and $\log g$ \citep{lan03}.  
We found that the equivalent width of the \ion{N}{3} $\lambda\lambda 
4510,4514,4518$ blend ($W_\lambda = 0.40 \pm 0.04$ \AA ) 
corresponds to a N abundance of $+0.68 \pm 0.09$ dex relative 
to the solar value.  The equivalent width of \ion{C}{3}
$\lambda 4187$ ($W_\lambda = 0.03 \pm 0.01$ \AA ) yields a
C abundance of $-0.55 \pm 0.20$ dex (i.e., sub-solar).  These trends 
are consistent with our expectations for CNO processed gas \citep{sch88}. 


\section{HST/STIS Spectroscopy and UV Wind Lines}   

The ultraviolet spectra of O-type stars are a key source of 
information about their photospheres and stellar winds 
\citep{wal85}, and we were fortunate to obtain the first ever 
UV spectra of LS~5039 with the {\it Hubble Space Telescope} 
Space Telescope Imaging Spectrograph (STIS) in 2002 August. 
During the first visit to the target (2002 Aug 13) we 
obtained two exposures (exposure times of 2087 and 1814 s)
with the G140L grating and one exposure (600 s) with the G230L
grating.  The G140L grating covers the wavelength range 
1150 -- 1730 \AA ~with a resolving power of 
$\lambda / \Delta\lambda = 1000$ while the 
G230L grating records the range 1570 -- 3180 \AA ~with
a resolving power of $\lambda / \Delta\lambda = 500$.
A second visit was made 7 days later (2002 Aug 20) to obtain a 
single exposure (2087 s) with the G140L grating.  
These visits correspond to orbital phases 0.145 and 0.708, 
respectively, when the O-star primary was oriented at 
an angle from the line of the nodes in the orbital plane of 
$\nu + \omega = 16^\circ$ and $121^\circ$, respectively (\S2). 
Thus, the first visit occurred just past the quadrature phase 
when the orbital axis was orthogonal to our line of sight and 
the second visit was made shortly after primary superior 
conjunction (compact companion in the foreground).  
There were no obvious flux or profile variations between 
these visits, so we formed an exposure weighted average 
FUV spectrum for examination.  

We obtained a single UV spectrum from the archives of 
the {\it International Ultraviolet Explorer 
Satellite}\footnote{http://archive.stsci.edu/iue} of the 
same O6.5~V((f)) reference standard shown above in 
Figure~2, HD~93146, to compare to that of LS~5039.  
We rebinned the high dispersion {\it IUE} spectrum 
(SWP~11136) of HD~93146 onto the wavelength grid of 
the STIS FUV spectrum, and both the reference and 
target spectra were rectified to a pseudo-continuum by 
division of a polynomial fit made through relatively 
line-free regions.   Two sections of these spectra
in the immediate vicinity of the strong 
stellar wind lines \ion{N}{5} $\lambda 1240$ 
and \ion{C}{4} $\lambda 1550$ are 
illustrated in Figures 4 and 5, respectively.
There is generally good agreement between the spectra 
of LS~5039 and HD~93146 in the UV photospheric lines. 
There are differences, however, in the interstellar line
strengths and especially in the wind lines.  
We find that the \ion{N}{5} $\lambda 1240$ P~Cygni profile 
has both a deeper blue absorption component (due to stellar 
wind material projected against the disk of the star) and 
a stronger red emission component (from scattered light surrounding 
the star) than is seen in the spectrum of HD~93146.  
The differences occur in the opposite sense in the 
\ion{C}{4} $\lambda 1550$ wind profile.   The enhanced 
N and weak C features mirror the results found in the optical 
spectra, and these same wind line differences are found in 
other ON type stars \citep{wal85,wal00}. 

\begin{figure}
\rotatebox{90}{
\epsscale{0.7}
\plotone{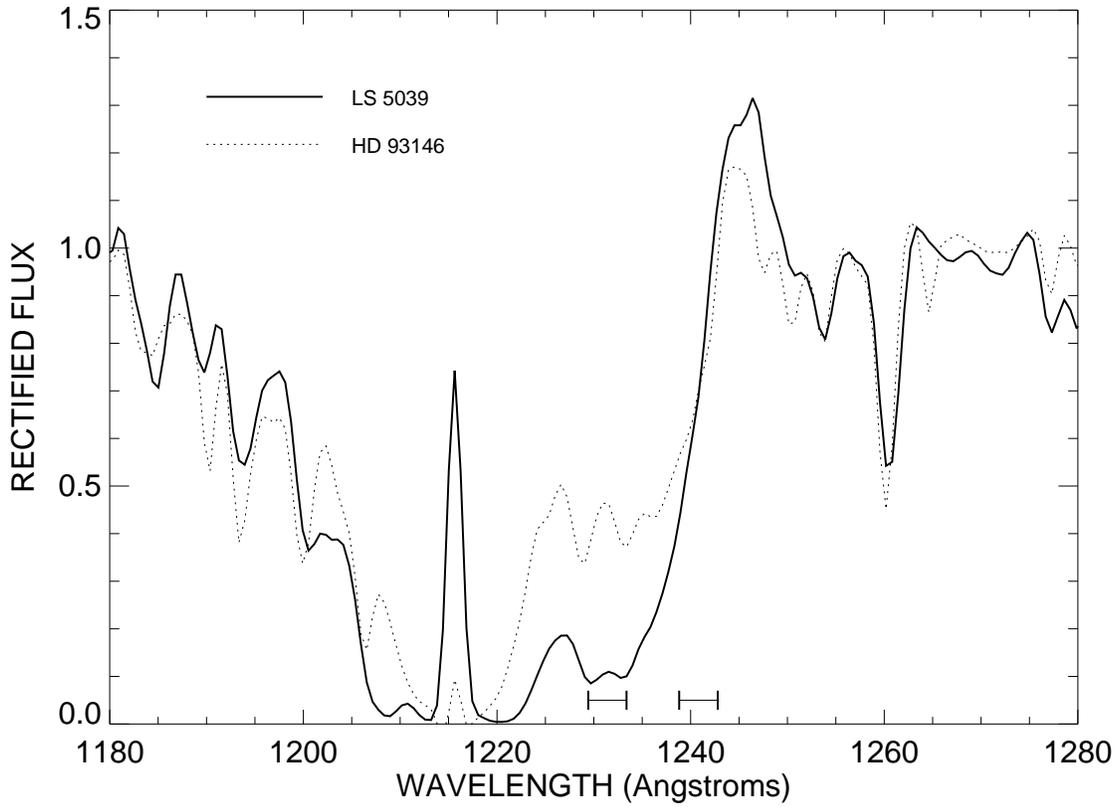}}
\caption{The \ion{N}{5} $\lambda\lambda 1238.821,1242.804$ 
P~Cygni profile in the spectrum of LS~5039 ({\it solid line}) 
compared to that of the O6.5~V((f)) standard star, HD~93146 
({\it dotted line}).   The error bars at the bottom show the 
rest wavelengths of the doublet ({\it right}) and the 
Doppler shifted wavelengths of the narrow absorption components
({\it left}).  The strong emission feature at 1216 \AA ~is 
geocoronal Ly$\alpha$.}
\label{fig4}
\end{figure}

\placefigure{fig4}     
  
\begin{figure}
\rotatebox{90}{
\epsscale{0.7}
\plotone{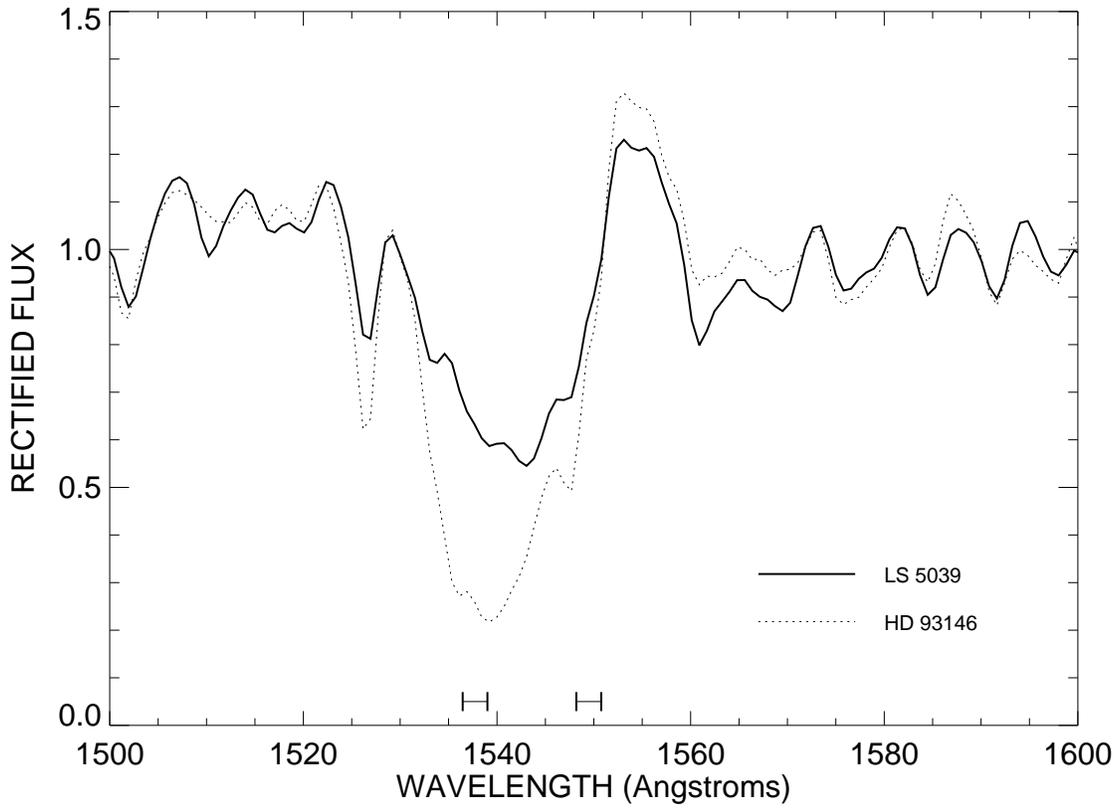}}
\caption{The \ion{C}{4} $\lambda\lambda 1548.195,1550.770$ 
P~Cygni profile in the spectrum of LS~5039 ({\it solid line}) 
compared to that of the O6.5~V((f)) standard star, HD~93146 
({\it dotted line}).   The error bars at the bottom show the 
rest wavelengths of the doublet ({\it right}) and the 
Doppler shifted wavelengths of the narrow absorption components
({\it left}).}
\label{fig5}
\end{figure}

\placefigure{fig5}     
  
Both wind profiles have shapes that are similar to their 
counterparts in the spectrum of HD~93146, and in particular, 
there are two minima in the absorption part of the 
\ion{N}{5} $\lambda 1240$ feature in the spectra of 
both stars.   These minima appear much narrower in the 
high dispersion version of the HD~93146 spectrum, and 
their spacing corresponds to the highly blueshifted 
positions of the \ion{N}{5} doublet.  They are examples 
of narrow (or discrete) absorption components (NAC) that often 
appear in the UV wind lines \citep{pri90,kap99}.  Detailed 
time sequences demonstrate that NAC usually proceed blueward 
through the trough and become narrower.  \citet{pri90} show 
that the sum $v_{\rm NAC} + {\rm HWHM}_{\rm NAC}$
is approximately constant and equal to the extreme velocity 
found in saturated line cores, and they argue that this 
quantity is a good estimate of the wind terminal velocity, 
$v_\infty$.   We measured the Doppler shifts of both 
minima in the \ion{N}{5} $\lambda 1240$ profile of LS~5039, 
and we find $v_{\rm NAC} = 2270 \pm 100$ km~s$^{-1}$. 
The resolving power of the STIS spectra is too low to determine the
NAC width, so we assumed it is the same as \citet{pri90} found
for HD~93146, ${\rm HWHM}_{\rm NAC} = 165$ km~s$^{-1}$. 
Thus, we estimate that the wind terminal velocity for LS~5039 
is $V_\infty = 2440 \pm 190$ km~s$^{-1}$, which agrees well 
with the mean for the O6.5~V subtype of $V_\infty = 2455$ km~s$^{-1}$
\citep{pri90}.  
  
In some MXRBs these wind lines display orbital phase variations 
due to excess ionization of the wind in a zone surrounding 
the X-ray source (the {\it Hatchett-McCray effect}; \citet{hat77,vlo01}).
The wind profiles of LS~5039 that we observed with HST/STIS showed 
no evidence of changes between the quadrature and superior conjunction 
phases.   The X-ray flux was moderately low at the time of 
the observations (see Fig.~7 below). 
If the X-ray source ionized a large volume that 
included regions projected against the stellar disk in 
the superior conjunction phase, then portions of the blue 
absorption trough would weaken as the scattering ions are 
ionized to higher levels \citep{vlo01}.   However, we 
suspect that the ionized region is small compared to the 
system dimensions in the case of LS~5039.
\citet{vlo01} use the same wind accretion model 
we discuss in \S6 to show that the $q$ parameter defining the 
extent of the ionization zone scales as $a^2 v_{\rm vel}^3$ 
among MXRBs, where $a$ is the separation of the components
and $v_{\rm vel}$ is the wind velocity relative to the neutron star. 
For the separation at the time of the 
{\it HST} near superior conjunction observation, we estimate that
this parameter falls in the range $q\approx 500 - 800$, 
based upon a comparison with results for Vela~X-1 from 
\citet{vlo01}.  This corresponds to a Str\"{o}mgren sphere 
radius of approximately $4\%$ of the system separation 
(see eq.~[4] in \citet{vlo01}).  The minimum projected separation between  
the outer edge of the ionization sphere and stellar limb at the observed phase 
is approximately $1.5 R_O$ (for an inclination $i=90^\circ$), so 
that no occultation is expected.  We suspect that any wind profile variations 
will be restricted to phases very close to superior conjunction 
and will only be observed if the system has a relatively large orbital 
inclination (\S6). 


\section{Reddening and Extinction}                  

The reddening and extinction of LS~5039 are directly related 
to the issues of the star's distance and space velocity (\S6). 
\citet{cla01} and \citet{rib02} both argue that the reddening 
is $E(B-V)=1.2\pm 0.1$, and \citet{rib02} suggest that the 
distance is $d=2.9\pm 0.3$~kpc based upon an assumed absolute 
magnitude.   The flux in the UV spectral range is very sensitive 
to the extinction curve \citep{fit99}, so we decided to 
re-investigate the reddening using the new STIS spectra.  
The STIS spectra are shown in Figure~6 in a version  
rebinned to a resolving power of $\lambda / \Delta\lambda = 100$.
We have included in Figure~6 fluxes based upon Johnson-Cousins and near-IR 
photometry from \citet{cla01} and the 2MASS survey \citep{cut03}, 
plus new Str\"{o}mgren $u, v, b$, and $y$ measurements we 
made on 2002 June 24 using the CTIO 0.9~m telescope and 
SITe $2048\times 2048$ CCD.  We also observed six
standard stars (HD~105498, HD~128726, HD~156623, HD~157795, HD~167321, and
HD~216743), taken from lists by \citet{cou87} and \citet{cla97},
in each band at a minimum of three different airmasses each to
calibrate the photometry.  All of the instrumental magnitudes were
determined using a large aperture of 8 arcsec, and the photometric
transformation equations from \citet{hen82} were applied to determine the
apparent magnitudes of LS~5039:
$u   =   13.25 \pm  0.05$, 
$v   =   12.72 \pm  0.04$,
$b   =   12.10 \pm  0.02$, and
$y   =   11.31 \pm  0.02$. 
The errors in the magnitudes are from the combined estimated instrumental 
errors and the errors in the transformation constants.

\begin{figure}
\rotatebox{90}{
\epsscale{0.7}
\plotone{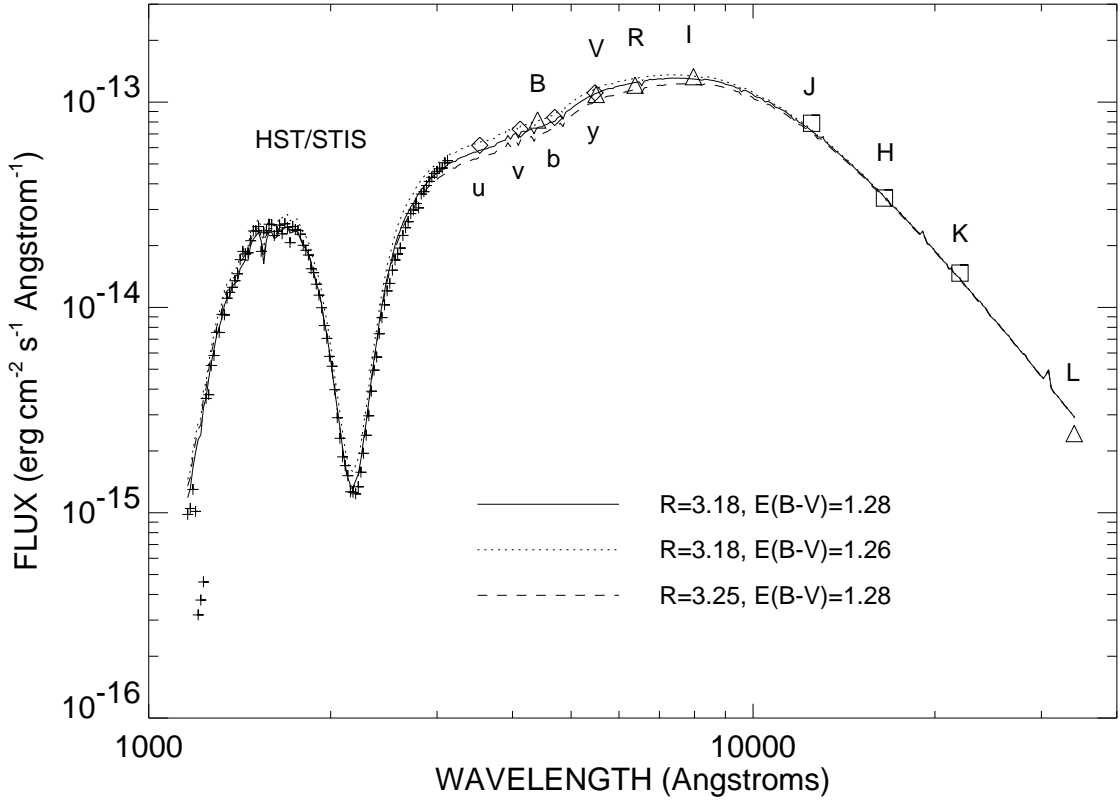}}
\caption{The UV, optical, and IR flux distribution of LS~5039
compared to model flux distributions for several 
assumptions about the ratio of total to selective extinction, $R$, 
and reddening, $E(B-V)$.  The data sources include 
UV fluxes from {\it HST}/STIS ({\it plus signs}),
CTIO Str\"{o}mgren photometry ({\it diamonds}),
Johnson $B, V$, Cousins $R, I$, and IR $L$-band photometry 
from \citet{cla01} ({\it triangles}), and 
IR $J, H, K$ photometry from the 2MASS survey \citep{cut03} 
({\it squares}).}
\label{fig6}
\end{figure}

\placefigure{fig6}     

The IR measurements were transformed to fluxes using the NICMOS units conversion 
tool\footnote{http://www.stsci.edu/hst/nicmos/tools/conversion\_form.html}. 
We determined fluxes from the optical band photometry using the IRAF
routine $calcphot$ in the {\it HST} {\it stsdas/synphot} package.  
This was done by comparing the difference in calculated magnitude 
between a reddened model spectrum of LS~5039 (a preliminary version 
of that discussed below) and a spectrum of Vega
with the observed difference in magnitude (using Vega magnitudes from 
\citet{bes83} and \citet{gra98}).  The model fluxes at the central 
filter wavelengths were then prorated to match the calculated and 
observed magnitude differences. 

We assumed that the model fluxes from \citet{lan03} 
for the spectroscopically determined parameters 
$T_{\rm eff} = 37.5$ kK and $\log g = 4.0$ provided a 
reliable estimate of the spectral flux distribution at 
the stellar photosphere.  We then calculated how the observed 
spectrum at Earth appears based upon the interstellar reddening 
curve given by \citet{fit99}.  There are three parameters that 
determine the appearance of the transformed spectrum: 
the reddening, $E(B-V)$, 
the ratio of total to selective extinction at $V$, 
$R = A(V) / E(B-V)$, and the ratio of stellar radius to distance, 
$R_O/d$ (equivalent to half the angular diameter in radians). 
The first two parameters determine the shape of the 
extinction curve, while the final parameter is a normalization factor. 
We formed $\chi^2$ residuals for a grid of $E(B-V)$ and $R$ values 
to find the best fit values that minimized $\chi^2$. 
The best fit, shown by the solid line in Figure~6, was made with 
$E(B-V)= 1.28 \pm 0.02$, $R= 3.18\pm 0.07$, and 
$R_O/d = (8.2\pm 0.6)\times 10^{-11}$ (angular diameter = 0.03 mas). 
This transformed spectrum fits well the entire FUV to IR range.  
We also show $1\sigma$ deviations from this 
fit in $E(B-V)$ ({\it dotted line}) and in $R$ ({\it dashed line}).

There are two nearby stars along the line of sight to LS~5039 
that have comparable extinction: 
BD$-14^\circ 5043$ (B1~IV) with $A_V=3.73$ and $d=1.1$~kpc 
\citep{kil93,he95} and
BD$-14^\circ 5040$ (O8~V) with $A_V=4.00$ and $d=1.9$~kpc 
\citep{kil93}.  Our derived extinction of $A_V=4.2 \pm 0.2$ 
indicates that the distance to LS~5039 is greater than 1.9~kpc, 
which is consistent with prior estimates ($d=2.9 \pm 0.3$~kpc;
\citet{rib02}).  
  

\section{Masses}                                    

In our previous paper \citep{mcs02}, we estimated the probable 
secondary mass for a range in primary mass by comparing the 
observed and predicted X-ray fluxes for wind accretion.  
Here we re-consider those arguments in light of the new 
results given above.  The independent parameter is the 
mass of the O star, $M_O$, that sets many of the other 
stellar parameters.  The gravity determination from \S3, 
$\log g = 4.0\pm0.1$, establishes the relationship between
mass and radius, $\log g_{\rm eff} = (1-\Gamma) G M / R^2$, 
where the ratio of radiative acceleration by electron scattering 
to gravitational acceleration is $\Gamma = 2.6 \times 10^{-5}
(L/L_\odot) / (M/M_\odot)$.  The specific relationship for 
LS~5039 based upon the derived effective temperature (\S3) is
$R_O/R_\odot = (1.56 \pm 0.16) (M_O/M_\odot)^{1/2}$.   
The fit of the observed flux distribution yields the angular 
diameter (\S5), which leads to a distance estimate of 
$d = (0.28 \pm 0.02) R_O/R_\odot$~kpc (somewhat closer than 
our prior estimate of $d = 0.32 R_O/R_\odot$~kpc; \citet{mcs02}). 

We estimate the wind mass loss rate using the equivalent width of
the H$\alpha$ emission line and the method of \citet{pul96}. 
The H$\alpha$ profile appears as a deep absorption line in 
stars with minimal wind, and the feature becomes progressively filled-in 
as the mass loss rate increases (appearing as a pure emission line 
in stars with strong winds).  
The mass loss rate depends upon the net emission equivalent 
width (i.e., the difference between the observed and purely 
photospheric absorption equivalent widths), the choice of the 
wind velocity law exponent $\beta$, the wind terminal velocity 
$v_\infty$, and the stellar effective temperature \citep{pul96}. 
We revised our assumptions about several of these parameters. 
We adopted $T_{\rm eff}=37.5$~kK (\S3) and used the detailed 
optical flux model from \citet{lan03} to find the appropriate 
photospheric equivalent width for H$\alpha$, $W_\lambda = 3.27$~\AA .
We adopted a velocity law index of $\beta = 1.0$ that \citet{pul96} 
advocate as the appropriate value for the weak wind case we find here. 
Finally, we set the terminal velocity to $V_\infty = 2440$ km~s$^{-1}$,
the value we estimated from the wind profiles in the STIS spectra (\S4). 
Most of these changes result in a moderate decrease in the estimated 
mass loss rate (Table~5, row 3) compared to our earlier results \citep{mcs02}. 

We noted in the previous paper that the H$\alpha$ strength appears
to be variable on time-scales longer than our observing runs \citep{mcs02}, 
which implies that there are long term variations in the mass loss rate. 
\citet{rei03} have recently shown that there are also long term variations 
in the X-ray flux that appear to be related to those observed in H$\alpha$. 
We show in Figure~7 the time evolution of the X-ray flux in the 0.3 -- 10 keV 
range reported by \citet{rei03}.  Figure~7 also illustrates the mass loss 
rate variations determined from the H$\alpha$ equivalent widths from 
\citet{mcs02} (3 measurements), \citet{rei03} (2 measurements), and our 
two recent observations ($W_\lambda = 2.83$~\AA ~and 2.95~\AA ~for 
2002 June 25 and 2003 March 21, respectively).   The $y$-axes in Figure~7 
were scaled to overlap the X-ray flux and mass loss estimates as closely as 
possible, and we confirm that these two quantities appear to vary in concert  
in the limited data available.   This result supports the wind accretion 
model and the validity of the mass loss estimates.  It also points out 
the need to compare the predicted and observed accretion luminosities 
at an epoch when nearly simultaneous optical and X-ray data are available. 
Here we will make the comparison for the closely spaced KPNO Coude Feed 
\citep{mcs02} and BeppoSAX \citep{rei03} observations made in 2000 October, 
which correspond to a mass loss rate and X-ray flux minimum in Figure~7.  

\begin{figure}
\rotatebox{90}{
\epsscale{0.7}
\plotone{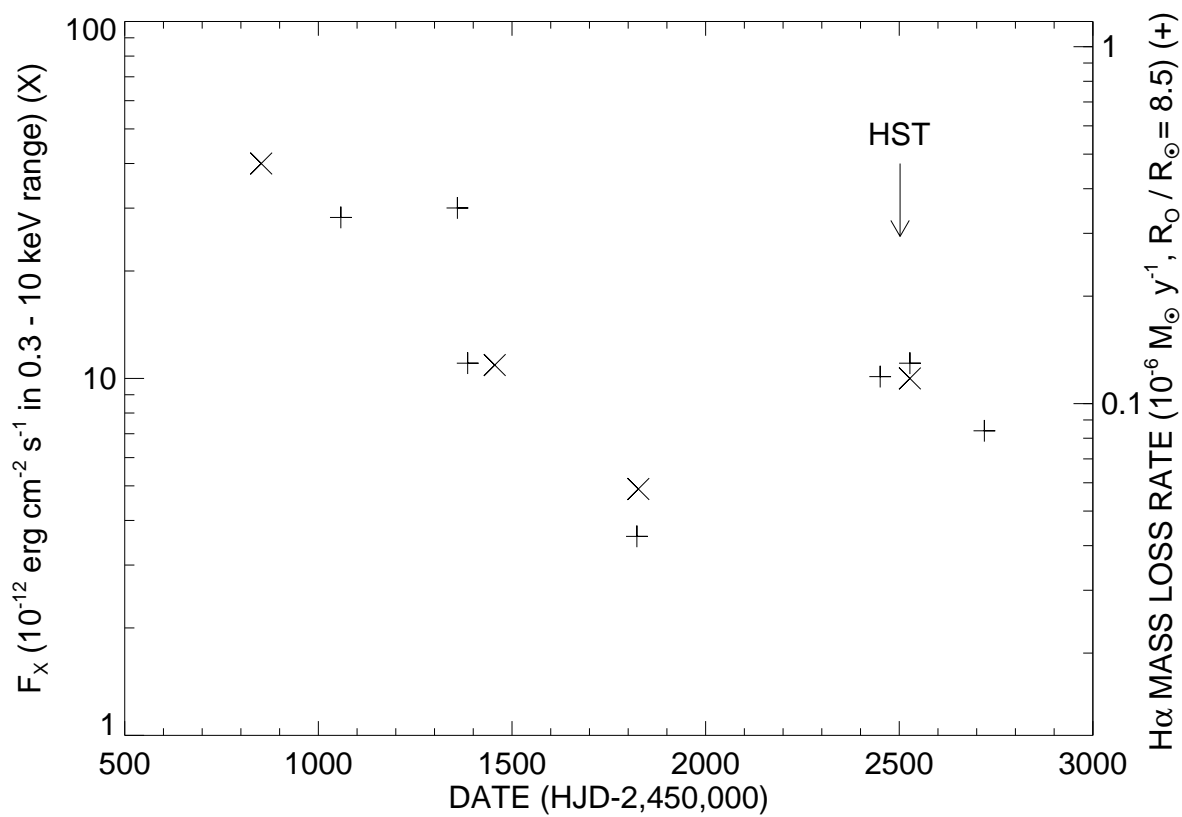}}
\caption{The co-variability of the observed X-ray flux in the 
0.3 -- 10 keV range \citep{rei03} ({X} {\it marks}) 
and H$\alpha$ mass loss rate ($+$ {\it signs}) as a function of time. 
The mass loss rate was determined assuming a stellar radius of 
$R_O/R_\odot = 8.5$, which corresponds to a mass of $M_O/M_\odot = 30$.
The date of the HST/STIS observation is indicated by an arrow.}
\label{fig7}
\end{figure}

\placefigure{fig7}     
  
We calculated the predicted wind accretion X-ray luminosity for a given 
$M_O$ and a grid of companion masses $M_X$ using the method of \citet{lam76}, 
and then we determined which $M_X$ value produced an X-ray luminosity
that matched the observed value.   This procedure was essentially the same 
as described earlier \citep{mcs02}, but we improved our accretion rate estimate 
by finding the mean of rates for equal time samplings around the elliptical 
orbit and by adopting the observed projected rotational velocity (\S3) 
in the calculation of the wind -- companion relative velocity (see eq.~[7b] 
in \citet{lam76}).   The resulting best-fit masses are listed in 
Table~5 and are illustrated in a mass plane diagram in Figure~8.   
Note that the stellar radius is well within its critical Roche radius 
at periastron in all these models, so that mass accretion by Roche lobe 
overflow is probably not a significant process in LS~5039. 
Our revised $M_X$ estimates are smaller than but comparable to what we found 
previously (see Fig.~2 in \citet{mcs02}).   The main reason for this difference 
is that we relied in the earlier paper on an X-ray luminosity estimate that 
turned out to be a local maximum (the first point shown in Fig.~7), and 
this consequently required a larger $M_X$ to account for the accretion luminosity.  

\placetable{tab5}      

\begin{figure}
\rotatebox{90}{
\epsscale{0.7}
\plotone{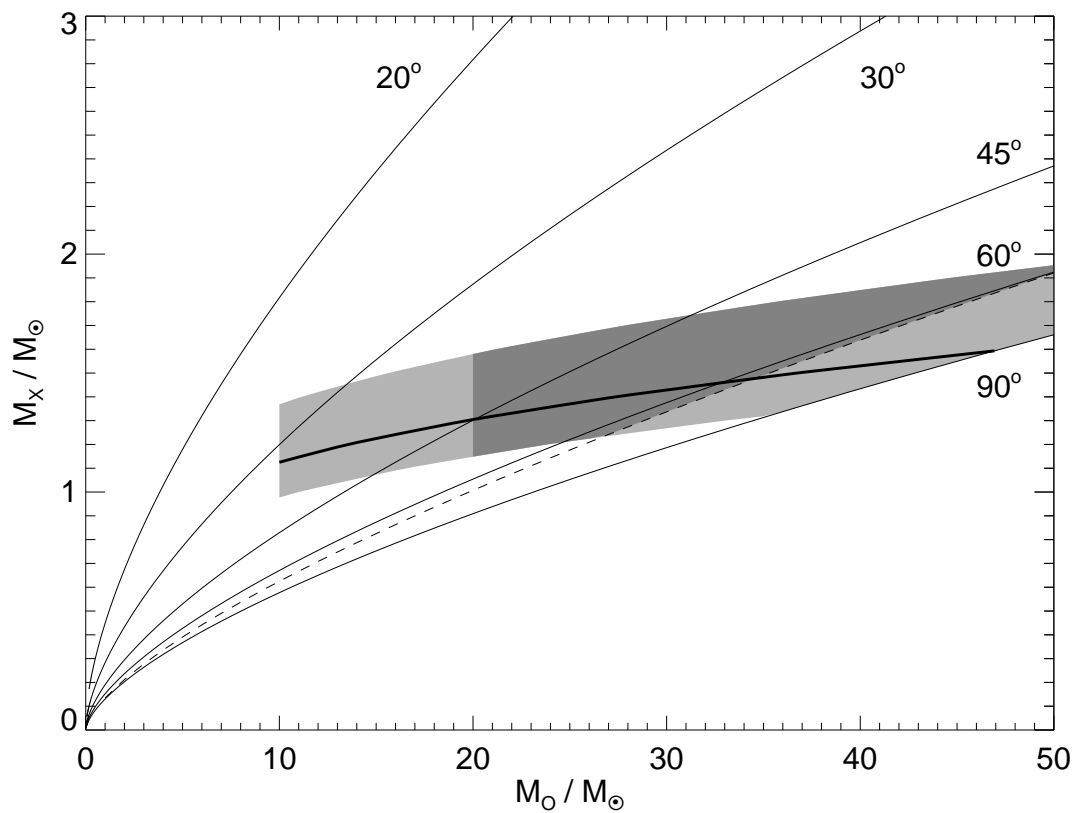}}
\caption{A mass plane diagram for LS~5039 with constraints from the
wind accretion model.  The thin solid lines trace the mass solutions 
from the orbital elements for the labelled values of orbital inclination.
The thick solid line shows the mass relationship derived by matching 
the observed and predicted X-ray fluxes from the wind accretion model 
(with the lightly shaded region showing possible solutions found by adjusting 
parameters discussed in the text).  The dashed line indicates the 
lower limit on $M_X$ established by the lack of observed eclipses, and 
the darker shaded region shows the most probable mass solution space.}
\label{fig8}
\end{figure}

\placefigure{fig8}     
  
The lightly shaded region surrounding the best-fit mass estimates in Figure~8 
shows how the solutions vary with changes in assumed parameters. 
The lower envelope corresponds to the calculations made using a 
wind velocity law exponent $\beta = 0.8$ instead of the 1.0 value
recommended by \citet{pul96}.  A lower exponent corresponds to a higher 
mass loss estimate (see Fig.~15 in \citet{pul96}), and this must be balanced 
by a lower mass accretor in order to match the observed X-ray luminosity 
(see eq.~[21] in \citet{lam76}).  The upper envelope shows the results 
obtained by assuming that the H$\alpha$ equivalent width was $1\sigma$
larger ($W_\lambda = 3.20$~\AA ) than actually observed in 2000 October 
($W_\lambda = 3.10$~\AA ).   The increased equivalent width corresponds 
to a weaker stellar wind (less emission), and in fact the assumed value is 
close to that for no wind ($W_\lambda = 3.27$~\AA ).  Here a reduction 
in wind strength is compensated by an increase in accretor mass in 
order to maintain the observed X-ray luminosity.  These tests indicate 
that our estimates for $M_X$ based upon the accretion model have 
formal errors of approximately $\pm 0.4 M_\odot$.  
However, we caution that the actual errors may be larger because the 
simple wind accretion model does not account for the effects of
the ionization of the wind by the X-ray source and because 
the efficiency parameter for conversion of accreted matter into X-ray 
radiation, $\zeta$, is only known approximately.   Nevertheless, 
our results for $M_X$ are consistent with a neutron star companion. 

There are several considerations that can help limit the 
range of acceptable mass for the O-star.  The observed reddening of 
LS~5039 compared to nearby, line-of-sight stars indicates a 
minimum distance of 1.9~kpc (\S5), and this distance corresponds to 
a lower limit on the stellar mass of $M_O > 20 M_\odot$.   
The wind accretion solutions at the high mass end require  
increased mass loss rates and larger system dimensions; 
such combinations occur at high inclinations according to the 
orbital mass function.   The dashed line in Figure~8 indicates the 
inclinations at which eclipses would begin to occur, and we see 
that for large $M_O$ the best-fit accretion models predict 
that eclipses should be observed.  However, there is no evidence of 
either X-ray \citep{rib99,rei03} or optical \citep{cla01} eclipses 
despite significant efforts to find them.  Thus, the probable absence of 
eclipses and the results of the wind accretion 
model suggest an upper mass limit of approximately $35 M_\odot$
(although masses as large as $50 M_\odot$ still fall within the 
error zone of the wind accretion model results). 

The wind properties offer additional constraints on the O-star mass. 
Hot massive stars have a wind momentum product, 
$\dot{M} ~v_\infty ~R_\star^{0.5}$, that is directly related to 
the stellar luminosity \citep{kud00}, and rows 3 and 4 of Table~5 
compare our H$\alpha$ derived mass loss rate with that predicted 
by the empirical wind momentum relationship for Galactic O-stars. 
The agreement appears best at the low mass end, but there is 
$\pm 1.2$~dex scatter in the wind momentum product among the stars 
that define the empirical relationship, so the full 20 -- $40 M_\odot$ 
range remains viable.  
Table~5 also lists the ratio $v_\infty / v_{\rm esc}$, which 
typically has a value of 2.6 in O-type stars \citep{lam99}. 
This criterion would again favor the lower mass range for LS~5039, 
$M_O \approx 20 M_\odot$, but we caution that there is a large spread in 
this ratio among hot stars.  Indeed, \citet{pri90} find a ratio
$v_\infty / v_{\rm esc} = 1.7$ for the comparable ON8~V star, HD~14633, 
and \citet{vlo01} quote ratio estimates in the range 0.8 -- 2.8 in other MXRBs. 

The final row of Table~5 gives estimates of the 
equatorial velocity $V_{\rm eq}$ that are derived from the 
projected rotational velocity assuming that the spin axis is 
aligned with the orbital axis (see Table~6 below) 
and that the orbital inclination can be set from  
the wind accretion models.  


\section{Pre-Supernova Orbital Parameters}          

The eccentricity of the system probably results from the 
supernova event that formed the neutron star, 
and the current value of the eccentricity can be used 
to infer the pre-SN orbital parameters, 
assuming that no asymmetrical kick velocity was imparted 
to the neutron star during the explosion and that the 
eccentricity has not decreased since the SN \citep{mcs02,rib02}. 
We show in the top portion of Table~6 estimates of 
the pre-SN orbital period, $P^{\rm initial}$, and 
mass of the SN-progenitor just prior to the SN, $M_2^{\rm initial}$, 
based upon the current orbital elements, mass solutions 
from the wind accretion model, and the assumption of $v_{\rm kick}=0$.    
There are several problems with these models that result from 
our revisions to the distance and eccentricity compared 
to earlier work \citep{mcs02}. 
The paramount difficulty is the apparent discrepancy between 
the runaway velocity expected from the eccentricity and masses, 
$v_{\rm sys}$~(predicted), and the peculiar space velocity, 
$v_{\rm pec}$~(derived) (based upon the observed proper motion 
and radial velocity, the distance associated with a given 
stellar mass $M_O$, and the method described in \citet{ber01}). 
The former is approximately twice as 
large as the latter, a much greater difference than expected 
from their formal errors (approximately 14\% and 19\%, respectively).  
The second problem is that the predicted pre-SN orbital period 
is so short that the Roche radius of the O-star is smaller than 
or equal to the estimated stellar radius.  A third, potential, 
difficulty is the relatively large mass of the SN-progenitor 
compared to theoretical expectations \citep{wel99}.  

\placetable{tab6}      

All of these problems can be alleviated by removing the 
assumption that the SN kick velocity was zero.  There is now 
a significant body of evidence that non-zero kick velocities 
must occur in SN \citep{bp95,vdh97,hug99,pfa02}.  
The introduction of a SN kick velocity will dramatically 
influence the post-SN orbital parameters and 
system runaway velocity.   Following the development of 
\citet{bp95}, we can express the final outcome parameters 
in terms of two ratios, 
\begin{displaymath}
m = {{M_2^{\rm initial} + M_O} \over {M_X + M_O}}
\end{displaymath}
and 
\begin{displaymath}
v = {v_{\rm kick} \over v_{\rm orb}},
\end{displaymath}
where $v_{\rm kick}$ and $v_{\rm orb}$ are the values of the 
kick velocity and pre-SN relative orbital velocity, respectively. 
We also need to specify the direction of the kick in terms of 
$\phi$, the angle between the SN-progenitor's orbital motion and 
the kick-velocity vector projected onto the pre-SN orbital plane, 
and $\theta$, the angle between the kick direction and the pre-SN orbital 
plane (see Fig.~1 in \citet{bp95}).   With these four parameters selected  
we can determine the post-SN eccentricity, system runaway velocity, and 
angle $\nu$ of misalignment between the O-star rotation axis and the current 
orbital angular momentum vector (see eq.~[2.8, 2.10, 2.16] in 
\citet{bp95}).  

We explored the possible solutions 
applicable to LS~5039 by calculating the predicted final eccentricity 
and runaway velocity over a grid of $m$, $v$, $\phi$, and $\theta$. 
Our strategy was to compute $e$ and $v_{\rm sys}$ for each set of 
parameters, and if the results in both cases were within $1 \sigma$ of 
the observed values, then we saved the increment of solid angle 
corresponding to the test values of $\phi$ and $\theta$ in our grid.
We omitted any models that led to a pre-SN orbit with an O-star 
Roche radius smaller than the observed value and any kick directions 
that would have the SN-remnant strike the O-star.  Our final results
are shown in Figure~9 for the masses associated with our $M_O = 30 M_\odot$
wind accretion model.  The grayscale intensity in this figure is directly 
proportional to the integrated solid angle of acceptable solutions for given 
values of $m$ and $v$, so that the figure may be interpreted as a
geometrical probability estimate (white corresponding to no solutions 
possible, black corresponding to the maximum integrated solid angle 
of acceptable solutions in the $(m, v)$ grid).   The high frequency
``ripples'' in the image result from the coarseness of our angular 
grid (increments of $5^\circ$ in $\phi$ and $\theta$) and our hard 
boundaries of $\pm 1 \sigma$ for acceptable solutions. 

\begin{figure}
\plotone{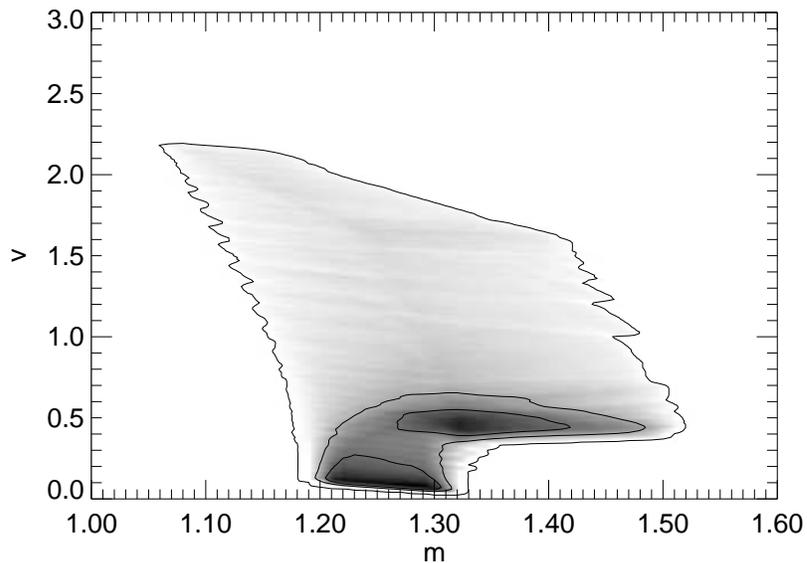}
\caption{A grayscale representation of the solution space of 
supernova mass loss parameters (assuming $M_O=30 M_\odot$ and 
$M_X=1.4 M_\odot$; the corresponding plots for the $M_O=20$ and $40 M_\odot$ cases
look similar except for small shifts in $m$).  The gray intensity indicates the 
fraction of solid angle into which a kick velocity is directed that 
results in an eccentricity and space velocity equal within errors 
to the derived values.  Contour lines are shown for acceptable 
solutions in 10, 5, and 0.1\% of the solid angle sphere 
(the darkest, maximum value corresponds to 20\%).    
The solutions are parameterized by 
$m$, the ratio of total binary mass before to that after the supernova, 
and $v$, the ratio of the kick velocity to the initial relative orbital
velocity \citep{bp95}.}
\label{fig9}
\end{figure}

\placefigure{fig9}     
  
There is a broad range of solutions in the $(m, v)$ plane, but 
the geometrically favored solutions are concentrated in two regions. 
The first occurs along a ridge in the lower portion of the diagram 
near $m=1.29$ and $v=0.08$.  The solutions correspond to kicks in 
the direction $\phi = 0^\circ - 55^\circ$, i.e., in approximately 
the same direction as the orbital motion of the progenitor at the 
time of the SN.   The pre-SN orbital parameters for such prograde 
($v_{\rm kick} +$) solutions are given in the middle portion of 
Table~6, which lists the 
solid angle weighted average values of initial period, kick velocity, and 
misalignment.  
We find initial periods similar to the no kick case 
but the progenitor masses are lower.  The associated kick   
velocities are small and consistent with the low values found by 
\citet{hug99} and by \citet{pfa02} for other MXRBs.  
\citet{pfa02} suggest that small kick velocities may occur in 
close binaries in which the SN-progenitor was a rapid rotator
at the time of the SN 
(i.e., in those which avoided spin down in a red-supergiant phase).
The other geometrically favored region occurs near 
$m=1.32$ and $v=0.46$, and these solutions correspond to retrograde kicks 
($v_{\rm kick} -$) with $\phi = 95^\circ - 180^\circ$.   
The lower portion of Table~6 gives the pre-SN orbital parameters 
for such solutions that have a longer initial period (and larger 
Roche radii) than the other solutions.  The kick velocities 
are much larger than before, but still within the range considered 
reasonable in other investigations \citep{bp95}.   However, the 
largest kick velocities permitted in Figure~9 (for $v_{\rm kick} \approx 1400$ 
km~s$^{-1}$ and $\nu > 90^\circ$) are probably too extreme for 
most of the commonly accepted kick velocity distributions.  

We expect that the binary probably suffered 
a significant reduction in orbital period during a spiral-in, 
common envelope phase prior to the SN explosion \citep{taa00}. 
At the conclusion of this stage the stars were in a 
short period orbit, and the strong tidal forces 
would ensure quick spin synchronization and circularization of 
the orbit \citep{zah77,cc97}.   We can estimate the average synchronous
rotation speed prior to the SN, $<V_{\rm sync}>$, using $R_O$ (Table~5) 
and $<P^{\rm initial}>$ (Table~6), and we list these estimates
for the $v_{\rm kick} +$ and $v_{\rm kick} -$ cases in Table~6.
The synchronous rotation speeds are smaller than the observed 
value of projected rotational velocity ($V\sin i = 140 \pm 8$ km~s$^{-1}$)
for the $v_{\rm kick} -$ models, so we prefer the 
solutions for the $v_{\rm kick} +$ models. 
The SN progenitor may also have been a rapid rotator as a result
of tidal synchronization, and the lower kick velocities associated 
with the $v_{\rm kick} +$ models are consistent with the suggestion 
that smaller kicks occur in rapid rotators \citep{pfa02}. 
However, regardless of the choice of the model for the SN event in LS~5039, 
it is clear that a modest, non-zero kick velocity is required to 
explain consistently the observed eccentricity and runaway velocity. 


\section{The Origin of the N Enrichment}            

It is somewhat surprising that we find clear evidence of CNO 
processed gas in the photosphere of LS~5039 while the  
spectral signature of CNO processed gas is generally absent in 
other MXRBs (with the interesting exceptions of 
Vela~X-1 = HD~77581 \citep{kap93} and 4U1700--37 = HD~153919 
\citep{cla02}).  The N-enriched gas has two 
possible sources, nuclear processed gas from the interior
of the O-star and/or processed gas from the envelope of the 
neutron star progenitor.   Here we briefly discuss the issues 
associated with both sources of enriched gas. 

The origin of the He and N enrichment in the OBN and other 
evolved massive stars is probably related to rotationally 
induced mixing of the stellar interiors.  
\citet{heg00} and \cite{mey00} have constructed stellar models 
that include rotation, and they find that significant mixing 
of CNO processed gas into the photosphere can occur over a main 
sequence lifetime in rapidly rotating, massive stars.  
\citet{how01} found that the OBN stars have a projected 
rotational velocity distribution consistent with the 
assumption that they represent a class of rapid rotators, 
and they showed that He is enhanced in several very rapidly 
rotating O-type stars.   

The projected rotational velocity of LS~5039 ($V\sin i = 140 \pm 8$ km~s$^{-1}$)
is not unusually large for O-type stars \citep{pen96,how97}, and the star's spin 
axis inclination, $i_{\rm spin}$, would need to be relatively small 
if the star is currently a rapid rotator (see $V_{\rm eq}$ estimates in Table~5).  
We showed above that the star could have been 
a rapid rotator due to tidal interactions prior to the SN 
(see the $v_{\rm kick} +$ solutions in Table~6).  
Nevertheless, we doubt that significant rotationally induced mixing 
has occurred since then because the elapsed time since the SN  
implied by the kinematical age ($<1$~My; \citet{rib02}) is much smaller 
than the characteristic time for large scale mixing (comparable to 
the main sequence lifetime, $\approx 6$~My; \citet{heg00,mey00}). 
Therefore, we doubt that the N-enrichment of LS~5039 is due
to internal mixing unless the O-star was born as a rapid 
rotator and mixing has progressed for a long time (i.e., 
the lifetime of the neutron star progenitor plus the kinematical age). 

The alternative explanation is that the nuclear processed gas 
was accreted from the SN progenitor.   The system had a 
short orbital period prior to the SN explosion (Table~6) 
that probably resulted from a spiral-in during a  
common-envelope phase (CEP) \citep{taa00}.   The surviving 
O-star probably accreted relatively little material 
during the CEP because of its short duration, and accretion 
following the CEP must have been very limited because 
mass transfer at that point would have caused an expansion of 
the orbit \citep{hil01}.  Instead we suspect that the CNO-enriched 
gas was transferred from the progenitor during a period 
immediately prior to the start of the CEP.   Massive donor
stars with radiative envelopes can in some circumstances sustain 
a stable mass transfer rate through Roche lobe overflow until 
the envelope is lost \citep{hje87,wel01,pod02}, and it is 
possible that several solar masses of enriched gas were  
accreted by the O-star gainer during this evolutionary stage. 
Other explanations such as direct wind fed accretion 
\citep{cla02} and accretion of the impacting SN shell \citep{fry81} 
are less compelling because the amount of enriched gas accreted in both 
cases is relatively minor. 

Many of the black hole MXRBs are also thought to 
have evolved from widely separated binaries into compact systems 
through a common envelope stage \citep{wel99,pod03}, and it is 
curious that these do not generally show the N enrichment we 
find in LS~5039.  The explanation may be related to the youth of 
LS~5039 \citep{rib02}.  We suspect that thermohaline mixing 
\citep{del94}, rotational mixing, and stellar wind mass loss 
will act to remove the enriched gas from the atmosphere so that 
the N enrichment may decline with the time elapsed since the SN.  
LS~5039 may represent an unusually 
young MXRB in which these processes have yet to 
make a significant change in the atmospheric abundances, 
so that the products of the intense binary 
interaction remain vividly intact in the stellar atmosphere.


\acknowledgments

We thank the staffs of CTIO, KPNO, and McDonald Observatory 
for their assistance in making these observations possible. 
We are especially grateful to Dr.\ Todd Henry and 
the SMARTS Consortium for the CTIO 1.5~m time awarded to this 
project.  We also thank Dr.\ Thierry Lanz and Dr.\ Ivan Hubeny for the  
use of and advice about their codes {\it Tlusty} and {\it Synspec}, 
Dr.\ Nolan Walborn for comments about the ON classification, 
Dr.\ Marc Rib\'{o} for communicating his radial velocity results
in advance of publication, and the referee 
Dr.\ Philipp Podsiadlowski for his insights on binary evolution. 
PJW is grateful for continuing hospitality at the
Department of Astrophysical Sciences, Princeton University.
Support for this work (HST Proposal Number 9449) was provided 
by NASA through a grant from the Space Telescope Science Institute,
which is operated by the Association of Universities for Research
in Astronomy, Incorporated, under NASA contract NAS5-26555.
Additional financial support was provided
by the National Science Foundation through grant AST$-$0205297 (DRG).
Institutional support has been provided from the GSU College
of Arts and Sciences and from the Research Program Enhancement
fund of the Board of Regents of the University System of Georgia,
administered through the GSU Office of the Vice President
for Research.  This research made use of
the Multimission Archive at the Space Telescope Science Institute (MAST),
NASA's Astrophysics Data System Bibliographic Service, and 
the Two Micron All Sky Survey, which is a joint project of the
University of Massachusetts and the Infrared Processing and Analysis 
Center/California Institute of Technology, funded by NASA and NSF.




\clearpage


\begin{deluxetable}{llrcccc}
\tablewidth{0pc}
\tablecaption{Spectrograph Parameters\label{tab1}}
\tablehead{
\colhead{UT Dates} &
\colhead{Instrument} &
\colhead{G/$\lambda_b$/O\tablenotemark{a}} &
\colhead{Filter} &
\colhead{CCD} &
\colhead{$\lambda / \Delta\lambda$} &
\colhead{Range (\AA)} }
\startdata
2002 Jun 07     \dotfill & McD 2.7 m/LCS & 1200/~4000/1 & \nodata  & TI    & ~2500 & 4060--4770 \\
2002 Jun 23--28 \dotfill & KPNO 0.9 m/CF &  316/12000/2 & OG550    & TI    & 11630 & 6532--6708 \\ 
2003 Mar 17     \dotfill & CTIO 1.5 m/CS &  831/~8000/2 & CuSO$_4$ & Loral & ~2320 & 4071--4741 \\
2003 Mar 18--20 \dotfill & CTIO 1.5 m/CS &  831/~8000/2 & BG39     & Loral & ~2320 & 4071--4741 \\
2003 Mar 21     \dotfill & CTIO 1.5 m/CS &  831/~8000/1 & GG495    & Loral & ~1800 & 5488--6799 \\
2003 Apr 26     \dotfill & McD 2.7 m/LCS & 1200/~4000/1 & \nodata  & Loral & ~2500 & 4071--4790 \\
2003 Apr 27--28 \dotfill & McD 2.7 m/LCS & 1200/~4000/1 & \nodata  & TI    & ~2500 & 4056--4764 \\
\enddata
\tablenotetext{a}{Grating grooves mm$^{-1}$/ blaze wavelength (\AA )/ order.}
\end{deluxetable}
\clearpage


\begin{deluxetable}{lcccl}
\small
\tablewidth{0pc}
\tablecaption{New Radial Velocities\label{tab2}}
\tablehead{
\colhead{HJD}           & 
\colhead{Orbital}       &
\colhead{$V_r$}         &
\colhead{$O-C$}         &
\colhead{}              \\
\colhead{(2450000$+$)}  &
\colhead{Phase}         &
\colhead{(km s$^{-1}$)} &
\colhead{(km s$^{-1}$)} &
\colhead{Source}}
\startdata
 2432.902 \dotfill &  0.901 & 
\phn         $ -17.8$ &\phn     $  -4.0$ & McD blue \\
 2448.858 \dotfill &  0.505 & 
\phn\phs     $  12.5$ &\phn\phs $   8.3$ & KPNO red \\
 2449.839 \dotfill &  0.727 & 
\phn\phn     $  -3.6$ &\phn\phs $   2.7$ & KPNO red \\
 2450.811 \dotfill &  0.946 & 
\phn         $ -11.7$ &\phn     $  -0.6$ & KPNO red \\
 2453.801 \dotfill &  0.622 & 
\phn\phn\phs $   5.9$ &\phn\phs $   7.1$ & KPNO red \\
 2715.894 \dotfill &  0.829 & 
\phn         $ -10.3$ &\phn\phs $   1.3$ & CTIO blue \\
 2716.902 \dotfill &  0.057 & 
\phn\phs     $  17.1$ &\phn     $  -1.4$ & CTIO blue \\
 2717.901 \dotfill &  0.282 & 
\phn\phs     $  17.0$ &\phn\phs $   2.6$ & CTIO blue \\
 2718.839 \dotfill &  0.495 & 
\phn\phn\phs $   9.5$ &\phn\phs $   4.9$ & CTIO blue \\
 2719.881 \dotfill &  0.730 & 
\phn         $ -17.2$ &         $ -10.7$ & CTIO red \\
 2755.901 \dotfill &  0.867 & 
\phn\phn     $  -9.9$ &\phn\phs $   3.3$ & McD blue \\
 2756.849 \dotfill &  0.081 & 
\phn\phs     $  28.6$ &\phn\phs $   7.8$ & McD blue \\
 2756.957 \dotfill &  0.106 & 
\phn\phs     $  31.7$ &\phs     $  10.3$ & McD blue \\
 2757.956 \dotfill &  0.331 & 
\phn\phn\phs $   6.9$ &\phn     $  -5.2$ & McD blue \\
\enddata
\tablecomments{See \citet{mcs01} for earlier measurements 
that were used in the orbital solution.}
\end{deluxetable}
\clearpage


\begin{deluxetable}{lc}
\tablewidth{0pc}
\tablecaption{Orbital Elements \label{tab3}}
\tablehead{
\colhead{Element}                 & \colhead{Value}  }
\startdata
$P$ (d)                  \dotfill & 4.4267    $\pm$ 0.0005  \\ 
$T$ (HJD - 2,450,000)    \dotfill & 2756.49   $\pm$ 0.07    \\ 
$K$ (km s$^{-1}$)        \dotfill & 17.6      $\pm$ 1.3     \\ 
$V_0$ (km s$^{-1}$)      \dotfill & 4.1       $\pm$ 0.8     \\ 
$e$                      \dotfill & 0.48      $\pm$ 0.06    \\ 
$\omega$ (deg)           \dotfill & 268       $\pm$ 10      \\ 
rms (km s$^{-1}$)        \dotfill &            5.5          \\ 
$f(m)$ ($M_\odot$)       \dotfill & 0.0017    $\pm$ 0.0005  \\ 
$a_1 \sin i$ ($R_\odot$) \dotfill & 1.36      $\pm$ 0.12    \\ 
\enddata
\end{deluxetable}
\clearpage


\begin{deluxetable}{lcc}
\tablewidth{0pc}
\tablecaption{He Line Ratios and Effective Temperature\label{tab4}}
\tablehead{
\colhead{Ratio} & 
\colhead{$\log [W_\lambda$(\ion{He}{2}) / $W_\lambda$(\ion{He}{1})$]$  } &
\colhead{$T_{\rm eff}$ (kK)}  }
\startdata
\ion{He}{2} $\lambda 4541$ : \ion{He}{1} $\lambda 4471$ \dotfill & $0.29\pm0.02$ & $37.9\pm0.4$ \\
\ion{He}{2} $\lambda 4541$ : \ion{He}{1} $\lambda 4387$ \dotfill & $0.68\pm0.03$ & $35.4\pm0.4$ \\
\ion{He}{2} $\lambda 4200$ : \ion{He}{1} $\lambda 4144$ \dotfill & $1.00\pm0.05$ & $36.3\pm0.3$ \\
\ion{He}{2} $\lambda 4200$ : \ion{He}{1} $\lambda 4713$ \dotfill & $0.86\pm0.03$ & $39.8\pm0.3$ \\
\enddata
\end{deluxetable}
\clearpage


\begin{deluxetable}{lccc}
\tablewidth{0pc}
\tablecaption{Range in Stellar Parameters\label{tab5}}
\tablehead{
\colhead{Parameter} &
\colhead{$M_O = 20 M_\odot$} &
\colhead{$M_O = 30 M_\odot$} &
\colhead{$M_O = 40 M_\odot$} }
\startdata
$R_O$ ($R_\odot$)  \dotfill                     & 7.0  & 8.5  & 9.9  \\
$v_\infty / v_{\rm esc}$ \dotfill               & 2.5  & 2.2  & 2.1  \\
$\log \dot{M}$ ($M_\odot$ y$^{-1}$)\tablenotemark{a}
                                       \dotfill &$-7.5$&$-7.4$&$-7.3$\\
$\log \dot{M}$ ($M_\odot$ y$^{-1}$)\tablenotemark{b}
                                       \dotfill &$-7.0$&$-6.8$&$-6.6$\\
$M_X$ ($M_\odot$) \dotfill                      & 1.3  & 1.4  & 1.5  \\
$R_O{\rm (Roche)}$ ($R_\odot$)\tablenotemark{c}
                               \dotfill         & 10.0 & 11.9 & 13.3 \\
$d$ (kpc) \dotfill                              & 1.9  & 2.4  & 2.7  \\
$V_{\rm eq}$ (km s$^{-1}$)\tablenotemark{d}\dotfill & 198  & 168  & 149  \\
\enddata
\tablenotetext{a}{From H$\alpha$ for the 2000 October minimum, the time of the 
KPNO Coude Feed \citep{mcs02} and BeppoSAX observations \citep{rei03}.}
\tablenotetext{b}{Based on the wind momentum relationship for Galactic
O-type stars of luminosity classes III and V \citep{kud00}.}
\tablenotetext{c}{Mean Roche radius of the O star at periastron.} 
\tablenotetext{d}{From $V\sin i$ and $i_{\rm spin}=i_{\rm orbit}$.}
\end{deluxetable}
\clearpage


\begin{deluxetable}{lcccc}
\tablewidth{0pc}
\tablecaption{Pre-Supernova Orbital Parameters\label{tab6}}
\tablehead{
\colhead{Parameter}          &
\colhead{$v_{\rm kick}$}     &
\colhead{$M_O = 20 M_\odot$} &
\colhead{$M_O = 30 M_\odot$} &
\colhead{$M_O = 40 M_\odot$} }
\startdata
$P^{\rm initial}$  (d) \dotfill                  & 0 & 1.39   & 1.39   & 1.39  \\
$M_2^{\rm initial}$ ($M_\odot$) \dotfill         & 0 & 11.6   & 16.6   & 21.5  \\
$v_{\rm sys}$ (predicted) (km s$^{-1}$) \dotfill & 0 & 181    & 210    & 233   \\
$v_{\rm pec}$ (derived) (km s$^{-1}$) \dotfill & \nodata & 95 & 118    & 137   \\
                                               &     &        &        &       \\  
$<P^{\rm initial}>$ (d) \dotfill               & $+$ & 1.52   & 1.58   & 1.63  \\
$M_2^{\rm initial}$ ($M_\odot$) \dotfill       & $+$ &  6.9   & 10.4   & 14.5  \\
$<v_{\rm kick}>$ (km s$^{-1}$) \dotfill        & $+$ & 58     & 51     & 41    \\
$<\nu >$ (deg) \dotfill                        & $+$ &  3     &  2     &  1    \\
$<V_{\rm sync}>$ (km s$^{-1}$)\dotfill         & $+$ & 232    & 273    & 306   \\
                                               &     &        &        &       \\  
$<P^{\rm initial}>$ (d) \dotfill               & $-$ & 4.61   & 4.60   & 4.81  \\
$M_2^{\rm initial}$ ($M_\odot$) \dotfill       & $-$ &  7.4   & 11.5   & 16.0  \\
$<v_{\rm kick}>$ (km s$^{-1}$) \dotfill        & $-$ & 182    & 211    & 226   \\
$<\nu >$ (deg) \dotfill                        & $-$ & 13     & 12     & 11    \\
$<V_{\rm sync}>$ (km s$^{-1}$)\dotfill         & $-$ & 77     & 94     & 104   \\
\enddata
\end{deluxetable}
\clearpage



\clearpage

\end{document}